## SHORT COMMUNICATIONS

# Dynamic van der Waals Interaction of a Moving Atom with the Walls of a Flat Slit

**G. V. Dedkov\* and A. A. Kyasov**

*Kabardino-Balkarian State University, ul. Chernyshevskogo 173, Nalchik, 360004 Kabardino-Balkarian Republic, Russia*
*\*e-mail: gv_dedkov@mail.ru*

**Abstract**—A general expression was obtained for the dynamic energy of the van der Waals interaction of a neutral atom with a flat slit whose walls are characterized by a frequency-dependent dielectric permittivity. The interaction of cesium atoms with the walls of metallic (Au) and dielectric (SiC) slits is analyzed numerically at speeds of $10^4$ to $10^7$ m/s. As the speed of atoms grows, the dynamic potentials near the walls become substantially smaller in magnitude than static potentials, but, in the intermediate region, the former exceed the latter by a factor of 1.5–2.0 in a specific range of speeds.

**DOI:** 10.1134/S1063784214040094

## INTRODUCTION

Advances in the realms of the synthesis and applications of nanostructures, such as fullerenes, nanotubes, and graphene shells, have quickened interest in problems associated with the interactions of charged particles and x rays with these and other materials as they propagate in nanoscopic channels and slits [1–3]. When neutral atoms and nanoparticles move in such channels, the interaction with the walls is due in the major part of the cross section to the van der Waals force, which, in addition to the velocity-independent component, develops dynamic components depending on the particle velocity. So far, researchers have not given much attention to those dynamic effects. In particular, experimental investigations devoted to the interactions of neutral atomic beams with flat surfaces and slits were performed only at atom speeds close to thermal values, in which case the dynamic effects are still insignificant, as will be shown below [4–6].

The objective of the present study is to calculate dynamic van der Waals potentials of interaction between an atom (nanoparticle) moving in a flat nanoscopic slit parallel to its walls with these walls.

## 1. THEORY

A general theory of the dynamic fluctuation-electromagnetic interaction (including van der Waals forces) of neutral atoms and nanoscopic particles with flat surfaces was developed in a series of our studies [7]. Following our method, we will consider a neutral spherical nanoparticle that has a temperature $T_1$ and a dipole polarizability $\alpha(\omega)$ and moving at a nonrelativistic speed $V$ in a flat vacuum slit between two infinitely long plates (parallel to them). Figure 1 gives a general view of the system being considered and basic geometric and physical parameters of the problem. The dielectric permittivities of the lower and upper plates, $\varepsilon_{1,2}(\omega)$, are assumed to be different in general, and their temperature is assumed to be $T_2$. We also assume fulfillment of the conditions

$$r_0 \ll \min(c/\omega_0, c/\omega_1, c/\omega_2),$$
$$r_0 \ll \min(z, l-z), \quad (1)$$

where $\omega_0$, $\omega_1$, and $\omega_2$ are characteristic frequencies in the absorption spectra of the particle and plates, $r_0$ is the particle radius, and $z$ is the distance from the particle to the lower plate (see Fig. 1). The conditions in (1) make it possible to treat the particle as a pointlike dipole and disregard the retardation of electromagnetic interactions. The disregard of the retardation effect is possible for the reason that, in the following,

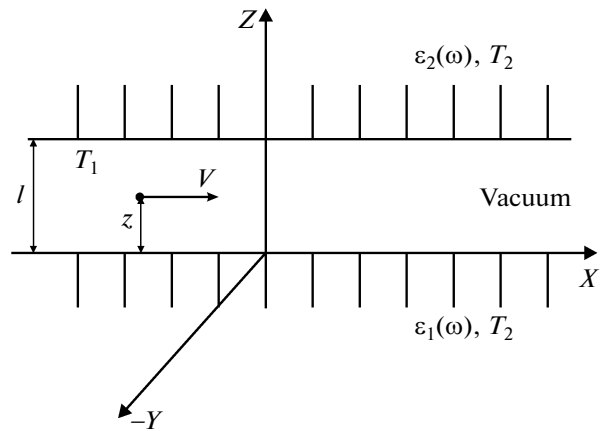

**Fig. 1.** System of coordinates and scheme of particle motion.

we consider nanoscopic slits of width not more than several tens of nanometers.

In the approximation being considered, the interaction of a moving particle with the fluctuation electromagnetic field of the slit is characterized by the normal (conservative) component

$$F_z = \langle \nabla_z (\mathbf{dE}) \rangle, \quad (2)$$

where $\mathbf{d} = \mathbf{d}^{sp} + \mathbf{d}^{in}$ is the total fluctuation particle dipole momentum consisting of the spontaneous and induced components, $\mathbf{E} = \mathbf{E}^{sp} + \mathbf{E}^{in}$ is the summed electric-field strength created within the slit by spontaneous and induced fluctuations, and angular brackets mean total quantum-statistical averaging. All of the quantities appearing in (2) are calculated on the basis of the formalism developed in [7] and in the reference frame associated with the plates at rest. The resulting formula for $F_z$ has the form [8]

$$F_z = -\frac{2\hbar}{\pi^2} \int\int\int_{0,\infty} d\omega\, dk_x\, dk_y\, k^2$$

$$\times \left\{ \mathrm{Re}\Delta_{12}^{(-)}(\omega)\mathrm{Im}\alpha(\omega + k_x V)\coth\frac{\hbar(\omega + k_x V)}{2k_B T_1} \right.$$

$$+ \mathrm{Re}\Delta_{12}^{(-)}(\omega)\mathrm{Im}\alpha(\omega - k_x V)\coth\frac{\hbar(\omega - k_x V)}{2k_B T_1} \quad (3)$$

$$+ \mathrm{Im}\Delta_{12}^{(-)}(\omega)\mathrm{Re}\alpha(\omega + k_x V)\coth\frac{\hbar\omega}{2k_B T_2}$$

$$\left. + \mathrm{Im}\Delta_{12}^{(-)}(\omega)\mathrm{Re}\alpha(\omega - k_x V)\coth\frac{\hbar\omega}{2k_B T_2} \right\},$$

$$\Delta_{12}^{(\pm)}(\omega) = \frac{\Delta_1(\omega)e^{-2kz} \pm \Delta_2(\omega)e^{-2k(l-z)}}{1 - \Delta_1(\omega)\Delta_2(\omega)e^{-2kl}}, \quad (4)$$

$$\Delta_1(\omega) = \frac{\varepsilon_1(\omega) - 1}{\varepsilon_1(\omega) + 1}, \quad \Delta_2(\omega) = \frac{\varepsilon_2(\omega) - 1}{\varepsilon_2(\omega) + 1}, \quad (5)$$

$$k = \sqrt{k_x^2 + k_y^2}.$$

Let us consider the particular case where the particle is at rest and where its temperature and the temperature of the plates are zero ($V = 0$ and $T_1 = T_2 = 0$). After some simple algebra in Eqs. (5), we obtain [8]

$$F_z = -\frac{2\hbar}{\pi}\int_0^\infty d\omega \int_0^\infty dk\, k^3 \alpha(i\omega)$$

$$\times \frac{\Delta_1(i\omega)e^{-2kz} - \Delta_2(i\omega)e^{-2k(l-z)}}{1 - \Delta_1(i\omega)\Delta_2(i\omega)e^{-2kl}}. \quad (6)$$

Taking into account the relation $F_z = -\partial U/\partial z$ between the force $F_z$ and the interaction energy (potential) $U(z)$, we can go over from expression (3) to an expression for $U(z)$. In particular, an expression for $U(z)$ in the case of identical materials of the plates was first obtained in [9]. Expression (3) is in perfect agreement with the result in [9] for this particular case. In the present study, however, we are interested in a dynamic generalization of expression (6) in the case of $V \neq 0$. For this, we go over in expression (3) to the limit of zero particle and slit-wall temperature:

$$\coth\left(\frac{\hbar(\omega \pm k_x V)}{k_B T_1}\right) \longrightarrow \mathrm{sgn}(\omega \pm k_x V),$$

$$\coth\left(\frac{\hbar\omega}{k_B T_2}\right) \longrightarrow \mathrm{sgn}(\omega). \quad (7)$$

With allowance for the relation $F_z = -\partial U/\partial z$, it follows from (3) and (7) that

$$U(V, z, l) = -\frac{\hbar}{2\pi^2}\int_{-\infty}^{+\infty} dk_x \int_{-\infty}^{+\infty} dk_y\, k$$

$$\times \mathrm{Im}\left[ i\int_0^\infty d\xi\, D(i\xi, z, l)\alpha(i\xi + k_x V)\right] \quad (8)$$

$$+ \frac{2\hbar}{\pi^2}\int_0^\infty dk_x \int_0^\infty dk_y\, k \int_0^{k_x V} d\omega\, D'(\omega, z, l)\alpha''(\omega - k_x V)$$

$$= u^{(0)} + \Delta U,$$

$$D(\omega, z, l)$$

$$= \frac{\Delta_1(\omega)\exp(-2kz) + \Delta_2(\omega)\exp(-2k(l-z))}{1 - \Delta_1(\omega)\Delta_2(\omega)\exp(-2kl)}, \quad (9)$$

$$\Delta_i(\omega) = \frac{\varepsilon_i(\omega) - 1}{\varepsilon_i(\omega) + 1}, \quad i = 1, 2. \quad (10)$$

In expression (8), one and two primes label, respectively, the real and imaginary parts of the functions involved; $l$ and $z$ are, respectively, the slit width and the distance from the atom to the left (for the sake of definiteness) wall of the slit; and the indices 1 and 2 number the slit walls, whose materials may have different dielectric permittivities. For $l \longrightarrow \infty$, expression (1) reduces to the respective expression for the potential of interaction of the atom with a single plate [10], while, at $V = 0$, it coincides with the expression presented in [8], since the contribution $\Delta U$ vanishes. As the speed of the atom grows, it is this term in expression (8) that becomes dominant. We indicate once again that expression (8) corresponds to the nonrelativistic dipole approximation of fluctuation electrodynamics.

Expanding the denominator on the right-hand side of (9) in a series, we obtain

$$D(\omega, z, l) = \sum_{n=0}^{\infty} [\Delta_1(\omega)^{n+1}\Delta_2(\omega)^n \exp(-2k(z + ln))$$
$$+ \Delta_1(\omega)^n \Delta_2(\omega)^{n+1} \exp(-2k(-z + ln + l))]. \quad (11)$$

Allowance in the sum in (1) for the zeroth term alone corresponds to the simplest additive approximation in which case the contributions from the two sides of the slit are summed independently. Substituting the expansion in (11) into expression (8), we can readily perform integration with respect to the wave vector $k_y$. For the ensuing calculations, we make use of the oscillator model of the atomic polarizability; that is,

$$\alpha(\omega) = \frac{\alpha(0)\omega_0^2}{\omega_0^2 - \omega^2 - i0\omega},$$
$$\alpha''(\omega) = \frac{\pi\alpha(0)\omega_0}{2}[\delta(\omega - \omega_0) - \delta(\omega + \omega_0)], \quad (12)$$

where $\omega_0$ is the characteristic frequency of atomic absorption and $\alpha(0)$ is the static polarizability. Upon the substitution of Eqs. (12) into the expression for $\Delta U$, integration with respect to the frequency can also be performed straightforwardly. Following this procedure and making the change of variables $\omega = \omega_0 t$, $k_x = x/2l$, and $k_y = y/2l$, we reduce the expressions for $U^{(0)}$ and $\Delta U$ to a form convenient for the ensuing numerical integration. Restricting ourselves for the sake of simplicity to the case of a homogeneous material of the slit walls, we arrive at

$$U^{(0)} = -\frac{\hbar\alpha(0)\omega_0}{4\pi^2 l^3}\sum_{n=0}^{\infty} Y_n(\lambda, \mu), \quad (13)$$
$$\lambda = V/2l\omega_0, \quad \mu = z/l,$$

$$Y_n(\lambda, \mu) = \int_0^{\infty}\int_0^{\infty} dx dt \frac{1 + t^2 - x^2\lambda^2}{(1 + t^2 - x^2\lambda^2) + 4x^2 t^2 \lambda^2}\Delta(t)^{2n+1} \quad (14)$$
$$\times [f(x(\mu + n)) + f(x(-\mu + n + 1))],$$

$$f(x) = K_0(x) + K_1(x)/x, \quad (15)$$

$$\Delta U = -\frac{\hbar\alpha(0)\omega_0}{8\pi l^3}\sum_{n=0}^{\infty}\int_{1/\lambda}^{\infty} dx x^2 \text{Re}[\Delta(x\lambda - 1)^{2n+1}] \quad (16)$$
$$\times [f(x(\mu + n)) + f(x(-\mu + n + 1))],$$

where $K_{0,1}(x)$ are Macdonald functions and $\Delta(t)$ is obtained from expression (10) upon the substitution $\omega = \omega_0 t$.

## 2. RESULTS OF NUMERICAL CALCULATIONS

In order to estimate the respective dynamic effect, we performed numerical calculations of the interaction of cesium atoms in the ground state $\{\alpha(0) = 57 \times 10^{-24}$ cm$^3$ and $\omega_0 = 1.44$ eV [11]$\}$ with the slit walls from gold and silicon carbide, approximating the dielectric functions for these materials in a standard form as

$$\varepsilon(\omega) = 1 - \frac{\omega_p^2}{\omega(\omega + i\gamma)},$$
$$\varepsilon(\omega) = \varepsilon_{\infty}\left(1 - \frac{\omega_p^2}{\omega(\omega + i\gamma)}\right). \quad (17)$$

The plasma frequency $\omega_p$, the damping parameter $\gamma$, and the high-frequency dielectric constants $\varepsilon_{\infty}$ for gold and silicon carbide were set to values corresponding to data from [12]. The results obtained by calculating $U(V, z, l)$ as a function of the relative coordinate $z/l$ for various values of the speed $V$ are shown in Figs. 2 (Au slit) and 3 (SiC slit). The values of $U(V, z, l)$ were normalized to the static-interaction energy [expression (8) at $V = 0$]. These figures show that the dynamic potentials differ substantially from their static counterparts. For a slit in a metal, the dynamic potentials become markedly weaker near the walls, but they may exceed the latter by a factor of 1.5 in intermediate channel parts when the speed $V$ exceeds $2 \times 10^6$ m/s. As the speed increases further, the potentials decrease, and only in the central part of the channel are they close to the static potentials. For a slit in SiC, the interaction potential undergoes more complicated changes. At atom speeds from $10^4$ to $2 \times 10^6$ m/s, the potential increases in the intermediate region, exceeding the static values by a factor greater than 2. As the speed increases further, the potential decreases over the entire channel width, becoming markedly smaller than the static values near the walls. For the sake of completeness, we indicate that, at distances from the wall that are about the interatomic spacing, the energy of the short-range (repulsive) interaction with the walls, which is disregarded in our calculations, must be added to the van der Waals potential. The inclusion of this repulsive interaction will lead to the reversal of the sign of the resulting interaction with the walls at short distances of $z < 0.4$ nm.

In conclusion, we note that, in the unique experiment reported in [5] and devoted to studying the propagation of a beam of sodium atoms within a flat slit formed by gold plates, the speeds of the atoms did not exceed $10^3$ m/s. In that case, the dynamic effects are quite small. Figures 2 and 3 show that, even at speeds around $10^5$ m/s (curves 1), the dynamic potentials differ insignificantly from their static counterparts over a major part of the channel cross section. Nevertheless, the dynamic decrease in the interaction at distances of $z/l < 0.1$ (in the case of a slit in a metal) becomes tenfold, and the effect in question may be noticeable at velocities of $10^4$ to $10^5$ m/s. A change in the interaction potential will manifest itself, for example, in the slit-width dependence of the intensity of the beam of neutral atoms (in the ground state) that was transmitted through the slit [5].

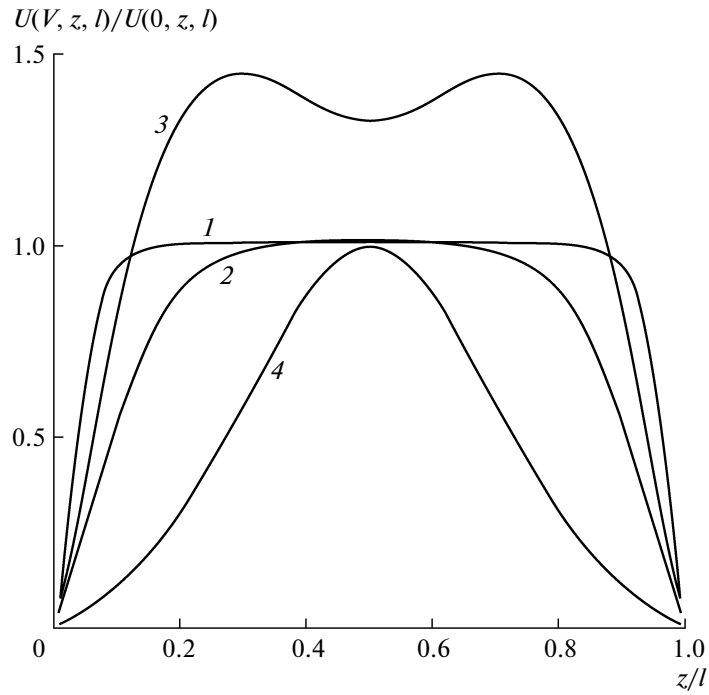

**Fig. 2.** Ratio of the dynamic to static van der Waals energy for a neutral cesium atom moving parallel to the walls of a slit in a metal (Au). Curves *1–4* correspond to the speeds of $V = 10^5$, $5 \times 10^5$, $2 \times 10^6$, and $4 \times 10^6$ m/s.

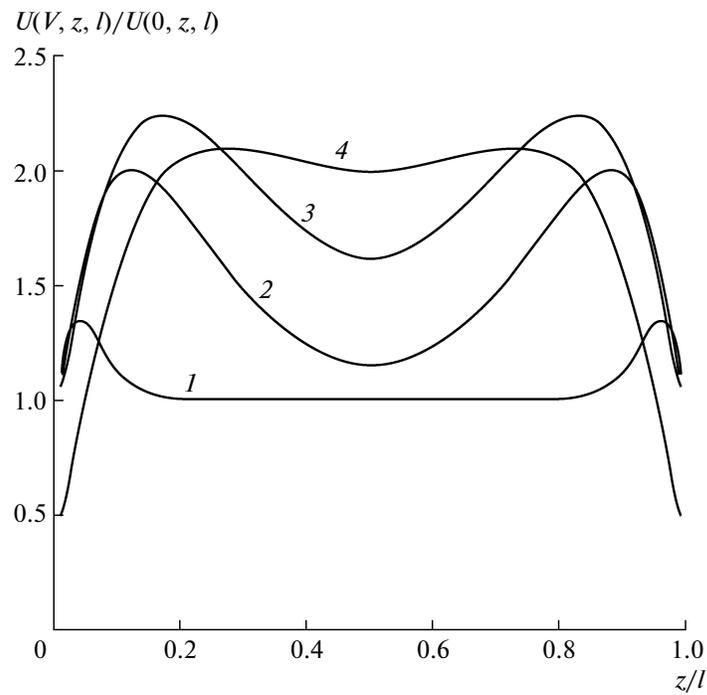

**Fig. 3.** As in Fig. 1, but for a slit in a SiC dielectric. Curves *1–4* correspond to the speeds of $V = 10^5$, $5 \times 10^5$, $1 \times 10^6$, and $4 \times 10^6$ m/s.

*Translated by A. Isaakyan*